\documentclass[review,number,sort&compress]{elsarticle}
\usepackage{lineno}

\usepackage{graphicx}
\usepackage{amssymb}
\usepackage{color,soul}
\usepackage{url}
\usepackage{ulem}
\usepackage[figuresright]{rotating}

\definecolor{orange}{rgb}{1,0.5,0}
\definecolor{cadmiumgreen}{rgb}{0.0, 0.42, 0.24}

\journal{Nuclear Instruments and Methods A}

\begin{document}

\begin{frontmatter}


\title{The light-yield response of a NE-213 liquid-scintillator detector
measured using 2 -- 6 MeV tagged neutrons} 


\author[lund,ess]{J.~Scherzinger}
\author[glasgow]{R.~Al~Jebali}
\author[glasgow]{J.R.M.~Annand}
\author[lund,ess]{K.G.~Fissum\corref{cor1}}
\ead{kevin.fissum@nuclear.lu.se}
\author[ess,midswe]{R.~Hall-Wilton}
\author[ess]{K.~Kanaki}
\author[m4]{M.~Lundin}
\author[ess,m4]{B.~Nilsson}
\author[lund,ess]{H.~Perrey}
\author[m4]{A.~Rosborg}
\author[m4,sweflo]{H.~Svensson}

\address[lund]{Division of Nuclear Physics, Lund University, SE-221 00 Lund, Sweden}
\address[ess]{Detector Group, European Spallation Source ERIC, SE-221 00 Lund, Sweden}
\address[glasgow]{SUPA School of Physics and Astronomy, University of Glasgow, Glasgow G12 8QQ, Scotland, UK}
\address[midswe]{Mid-Sweden University, SE-851 70 Sundsvall, Sweden}
\address[m4]{MAX IV Laboratory, Lund University, SE-221 00 Lund, Sweden}
\address[sweflo]{Sweflo Engineering, SE-275 63 Blentarp, Sweden}

\cortext[cor1]{Corresponding author. Telephone:  +46 46 222 9677; Fax:  +46 46 222 4709}
\fntext[fn2]{present address: University of Glasgow, Glasgow G12 8QQ, Scotland, UK}

\begin{abstract}
The response of a NE-213 liquid-scintillator detector has been measured 
using tagged neutrons from 2--6 MeV originating from an Am/Be neutron 
source. The neutron energies were determined using the time-of-flight
technique. Pulse-shape discrimination was employed to discern between 
gamma-rays and neutrons. The behavior of both the fast (35~ns) and the 
combined fast and slow (475~ns) components of the neutron 
scintillation-light pulses were studied. Three different prescriptions 
were used to relate the neutron maximum energy-transfer edges to the 
corresponding recoil-proton scintillation-light yields, and the results 
were compared to simulations. The overall normalizations of 
parametrizations which predict the fast or total light yield of the 
scintillation pulses were also tested. Our results agree with both 
existing data and existing parametrizations. We observe a clear 
sensitivity to the portion and length of the neutron scintillation-light 
pulse considered.
\end{abstract}

\begin{keyword}
NE-213, gamma-rays, fast-neutrons, scintillations, pulse-shape discrimination
\end{keyword}

\end{frontmatter}

\section{Introduction}
\label{section:introduction}

Organic liquid scintillators are typically employed to detect fast 
neutrons in mixed neutron and gamma-ray fields.  When exposed to 
these two different types of radiation, such scintillators emit light 
with dramatically different decay-time constants. Gamma-rays interact 
dominantly with the atomic electrons of the scintillator atoms. The 
freed electrons are almost minimum ionizing and produce very fast 
flashes of light (10s of ns decay times). In contrast, neutrons 
interact dominantly with the hydrogen nuclei (and to a lesser extent, 
carbon nuclei) of the scintillator molecules via scattering. Only the 
hydrogen nuclei are given sufficient energy to produce a significant 
signal, and in the neutron energy range from 2--6~MeV, the recoiling 
protons are far from minimum ionizing and produce much slower flashes 
of light (100s of ns decay times). By examining the time dependence of 
the scintillation-light intensity, differences in pulse shapes may be 
identified. Such pulse-shape discrimination (PSD) may be used to 
determine whether or not the incident radiation was a neutron or 
gamma-ray.

The organic liquid scintillator NE-213~\cite{ne213} has been a popular 
detector medium since its introduction in the early 
1960s~\cite{batchelor61}. It is a solution of aromatic molecules 
suspended in a xylene solvent\footnote{More recent variants are based 
on a pseudo-cumene solvent. These variants are less flammable than the 
xylene-based original.}. The result is a flammable, corrosive, 
sharp-smelling liquid with a flash point of $\sim$26~$^{\circ}$C that 
poses a considerable health risk. Nevertheless, due to its strong 
gamma-ray rejection properties, which are facilitated by excellent 
PSD characteristics and high detection efficiency for fast neutrons, 
NE-213 (first three scintillation-light decay constants: 3.16, 32.3, 
and 270~ns) has long set the standard for organic liquid scintillators 
(and beyond).  As a result, newly developed fast-neutron detectors are 
often compared to it~\cite{bayat12,iwanowska13,pawelczak13,jebali15}.

We have recently reported on a technique for tagging neutrons emitted 
by actinide/Be-based radioactive sources~\cite{scherzinger15}. In that 
paper, we also discuss in detail the experimental apparatus and 
technique employed here. In this paper, we present the results of our 
inaugural investigation performed using this neutron-tagging 
technique -- a precision mapping of the response of a NE-213 
liquid-scintillator detector using neutrons tagged from 2--6 MeV.

\section{Apparatus}
\label{section:apparatus}

\subsection{Actinide/Be-based source}
\label{subsection:be_source}

An 18.5 GBq $^{241}$Am/$^{9}$Be (Am/Be) source was employed for the 
irradiations performed in this work. We note that the neutron-tagging
technique described below will work equally well for any 
actinide/Be-based neutron source. Unwanted 60~keV gamma-rays 
associated with the $\alpha$-decay of $^{241}$Am were attenuated using 
a 3~mm thick Pb sheet. The source radiated 
(1.106$\pm$0.015)~$\times$~10$^6$ neutrons per second nearly 
isotropically~\cite{{natphyslab}}. Fast neutrons were produced when 
the $\alpha$-particles from the decay of $^{241}$Am interacted with 
the $^{9}$Be. These neutrons had a maximum energy of about about 
11~MeV~\cite{lorch73}. Roughly 25\% of the neutrons had an energy less 
than 1~MeV~\cite{vijaya73}. If the recoiling $^{12}$C was left in its 
first excited state (about 55\% of the 
time~\cite{vijaya73,mowlavi04,liu07}), the freed neutron was 
accompanied by an isotropically radiated prompt 4.44~MeV de-excitation 
gamma-ray. The half-value layer (HVL) for 3.5~MeV gamma-rays in lead is 
1.51~cm, and above this energy, the HVL does not increase with 
increasing gamma-ray energy. As a result, fewer than 20\% of these 
4.44~MeV gamma-rays were attenuated in the 3~mm Pb sleeve. Thus, the 
radiation field associated with the lead-shielded Am/Be source was to 
a large extent a combination of 4.44~MeV gamma-rays and their 
associated fast-neutrons.

\subsection{NE-213 liquid-scintillator detector}
\label{subsection:ne213_detector}

\begin{figure} 
\begin{center}
\resizebox{0.60\textwidth}{!}{\includegraphics{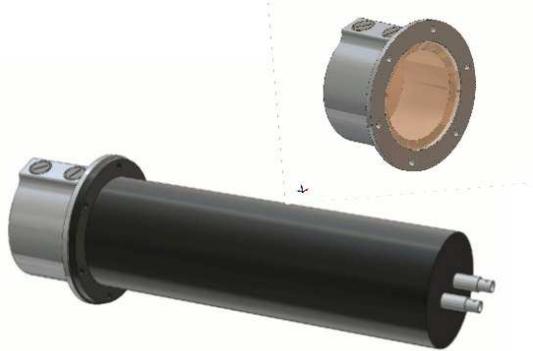}}
\caption{\label{figure:ne213_detector}
The NE-213 detector. Top: the scintillator ``cup". The optical 
boundary is provided by a borosilicate-glass window (light brown). 
Bottom: The gray cylinder to the left is the ``cup". The black 
cylinder to the right is the $\mu$-metal shielded PMT and base 
assembly. Figure from Ref.~\cite{jebali15}. (For interpretation of 
the references to color in this figure caption, the reader is 
referred to the web version of this article.)
}
\end{center}
\end{figure}

Figure~\ref{figure:ne213_detector} presents the NE-213 
liquid-scintillator detector employed in this measurement. A 3~mm 
thick cylindrical aluminum cell 62~mm deep by 94~mm in diameter, 
coated internally with EJ-520 TiO$_2$-based reflective 
paint~\cite{ej520}, contained the NE-213. A 5~mm thick borosilicate 
glass plate~\cite{borosilicate}, attached using Araldite 2000$+$ 
glue~\cite{araldite}, served as an optical window. A pressurized 
nitrogen gas-transfer system was used to fill the cell with 
nitrogen-flushed NE-213, and Viton O-rings~\cite{viton} were used 
to seal the filling penetrations. The filled cell was coupled to a 
cylindrical PMMA UVT lightguide~\cite{pmma} 57~mm long by 72.5~mm 
in diameter coated on the outside by EJ-510~\cite{ej510} 
TiO$_2$-based reflector. The cell/lightguide assembly was attached 
to a spring-loaded, $\mu$-metal shielded 3~in. ET Enterprises 9821KB 
photomultiplier tube (PMT) and base~\cite{et_9821kb}. Gain for the 
NE-213 detector was set using standard gamma-ray sources, resulting 
in an operating voltage of about $-$2000 V. Typical signal risetime 
was 5~ns. 

\subsection{YAP(Ce) 4.44~MeV gamma-ray detectors}
\label{subsection:yaps}

\begin{figure} 
\begin{center}
\rotatebox{90}{
\resizebox{0.40\textwidth}{!}{\includegraphics{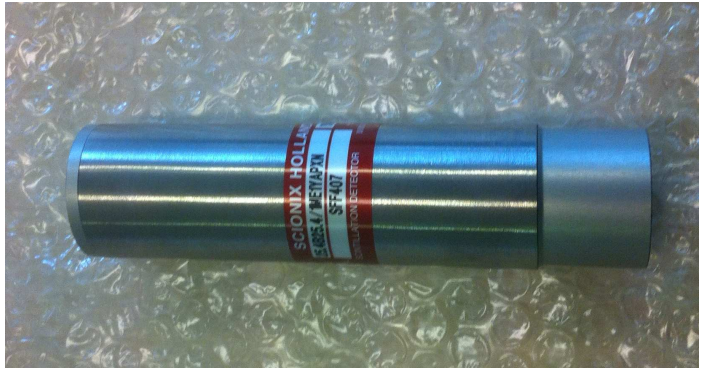}}
}
\caption{\label{figure:yap_detector}
Photograph of a YAP(Ce) detector. A cylindrical 1~in. long by 1~in. 
in diameter crystal (right) was coupled to 1~in. diameter PMT. 
Figure from Ref.~\cite{scherzinger15}.
}
\end{center}
\end{figure}

Figure~\ref{figure:yap_detector} presents a photograph of one of 
the YAP(Ce) gamma-ray detectors provided by Scionix~\cite{scionix} 
that was employed in this measurement. YAP(Ce) is an abbreviation 
for yttrium aluminum perovskit:cerium, or YAlO$_{3}$, Ce$^{+}$ 
doped. YAP(Ce) is both radiation hard and relatively insensitive 
to fast neutrons. Each detector was composed of a cylindrical 
1~in. long by 1~in. diameter crystal~\cite{moszynski98} attached 
to a 1~in. Hamamatsu Type R1924 PMT~\cite{hamamatsu}. Gains for 
the YAP(Ce) detectors were set using standard gamma-ray sources 
with typical operating voltages of about $-$800~V. Typical signal 
risetime was 5~ns. The energy resolution for the 662~keV peak of 
$^{137}$Cs measured using such a detector was about 10\%. We stress 
that the YAP(Ce) detectors were not used for gamma-ray spectroscopy, 
but rather to count the 4.44~MeV gamma-rays emitted by the source 
and thus provide a reference in time for the corresponding emitted
neutrons.

\section{Measurement}
\label{section:measurement}

\subsection{Setup}
\label{subsection:setup}

\begin{figure} 
\begin{center}
\resizebox{12.5cm}{9cm}{\hspace{-1.5cm}\includegraphics{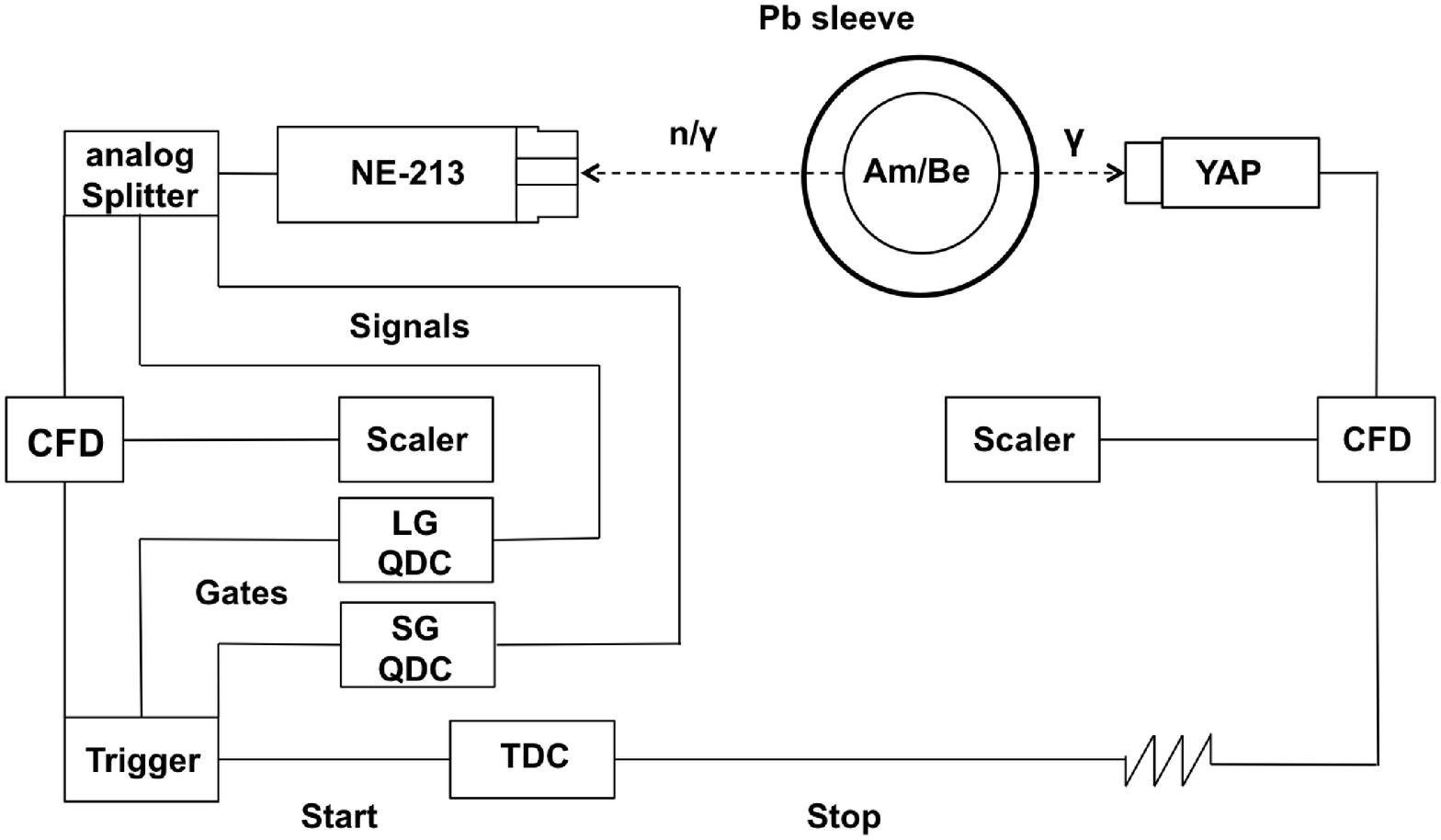}}
\caption{\label{figure:configuration}
A simplified schematic of the experimental setup. The Am/Be source, 
the Pb sleeve, a single YAP(Ce) detector, and the NE-213 detector 
are all shown together with a block electronics diagram. Figure 
from Ref.~\cite{scherzinger15}.
}
\end{center}
\end{figure}

Figure~\ref{figure:configuration} shows a simplified block diagram 
of the experimental setup. As previously mentioned, the Am/Be 
source was placed within a 3~mm thick Pb sleeve to attenuate the 
source-associated 60~keV gamma-rays. Two YAP(Ce) detectors (for 
simplicity, only one is shown) were located about 5~cm from the 
Am/Be source at source height. The threshold for the YAP(Ce) 
detectors was about 350~keV$_{ee}$ (keV electron equivalent). The 
NE-213 detector was located 2.420~m from the Am/Be source and also 
at source height. The threshold for the NE-213 detector was about 
250~keV$_{ee}$. Both detectors triggered overwhelmingly on the 
source-associated 4.44~MeV gamma-rays corresponding to the decay 
of the first excited state of $^{12}$C, but they also registered a 
large number of 2.23~MeV gamma-rays associated with neutron capture 
on the hydrogen in the water and paraffin used as general radiation 
shielding (not shown in this simplified illustration). The NE-213 
detector also triggered on the fast neutrons radiated from the 
source. By detecting both the fast neutron and the prompt correlated 
4.44~MeV gamma-ray, neutron time-of-flight (TOF) and thus energy was 
determined. This neutron-tagging technique enabled the mapping of 
the response of the NE-213 cell to fast neutrons as a function of 
their kinetic energy. Note that due to the energy invested in the 
4.44~MeV gamma-ray, the tagging technique restricted the maximum 
available tagged-neutron energies to about 6~MeV.

\subsection{Electronics and data acquisition}
\label{subsection:electronics_and_daq}

The analog signals from the detectors were split and sent to LeCroy 
2249A (DC-coupled short gate SG) and 2249W (AC-coupled long gate LG) 
CAMAC charge-to-digital converters (QDCs) and Phillips Scientific 
715 NIM constant-fraction timing discriminators (CFDs). The 
discriminator logic signals were sent to LeCroy 4434 scalers and 
LeCroy 2228A CAMAC time-to-digital converters (TDCs). A CES 8210
branch driver was employed to connect the \mbox{CAMAC} electronics 
to a VMEbus and a SBS 616 PCI-VME bus adapter was used to connect
the VMEbus to a LINUX PC-based data-acquisition (DAQ) system. The 
signals were recorded and processed using ROOT-based 
software~\cite{root}. Signals from the NE-213 detector were used to 
trigger the DAQ and also provided the start for the TOF TDC. As 
previously mentioned, the NE-213 detector QDCs included a 60~ns 
gated SG QDC and a 500~ns gated LG QDC, where in both cases, the 
gates opened 25~ns prior to the arrival of the analog pulse. The 
YAP(Ce) detector provided the stop signal for the TOF TDC. We were 
particularly interested in two source-related event types: 1) a 
fast neutron detected in the NE-213 detector (which started the 
TOF TDC) and the corresponding 4.44~MeV gamma-ray detected in the 
YAP(Ce) detector (which stopped the TOF TDC); and 2) a prompt, 
time-correlated gamma-ray pair detected one in the NE-213 detector 
and one in a YAP(Ce) detector (a gamma-flash event, see below).
Such a pair of gamma-rays can result from, for example, the 
$\alpha$ decay of $^{241}$Am to the higher excited states of 
$^{237}$Np.
 
\subsection{Energy calibration}
\label{subsection:energy_calibration}

Gamma-ray sources are typically used to calibrate organic 
scintillators as the light yield of the recoiling atomic electrons 
is linear above about 100~keV~\cite{knox72,knoll89}. However, the 
low Z value typical of liquid scintillators means that gamma-ray 
interactions are dominated by Compton scattering at energies of a 
few MeV. Thus, resolution-broadened Compton edges must be carefully 
interpreted in order to calibrate the detector. Two different 
prescriptions to extract the Compton edge from a resolution-smeared 
distribution have been reported by Flynn~et~al.~\cite{flynn64} and 
Knox and Miller~\cite{knox72}. More recently, with the aid of Monte 
Carlo simulations, it has become generally accepted that the Compton 
edge lies somewhere between these 
prescriptions~\cite{beghian65,dietze82,arneodo98,matei12}. 
Apparently, no clear concensus exists.

We simulated the response of our detector to gamma-rays (and then 
neutrons, see below) using GEANT4 (version 10.00 patch2) with the 
standard electromagnetic-interaction package and hadronic physics 
list QGSP\_BERT\_HP which provided high-precision data-driven models 
for neutron interactions below 20~MeV~\cite{agostinelli03,allison06}. 
The amplitude of the detector signal was provided by a 
sensitive-detector class which recorded the total energy deposited 
in the liquid-scintillator volume. The detector was defined to be 
the NE-213 filled cell together with the non-sensitive PMMA 
lightguide. For the purpose of the energy-calibration simulation, a 
point source of gamma-rays was positioned along the cylindrical 
symmetry axis of the cell at a distance of 1.5~cm from the face. The 
gamma-rays were directed onto the cell along its symmetry axis. 
Simulations of the deposited energy / scintillation-light yield for 
the detector were performed for the gamma-rays coming from $^{22}$Na 
(511~keV and 1274~keV) and $^{137}$Cs (662~keV), with corresponding 
Compton-edge-equivalent energies of 341~keV$_{ee}$, 1062~keV$_{ee}$, 
and 477~keV$_{ee}$, respectively. A non-linear, energy-dependent 
parametrization of the detector resolution measured for gamma-ray 
energies between 0.5~MeV$_{ee}$ (18\%) and 4.0~MeV$_{ee}$ (10\%) 
was included in the simulation. Note that this exact same 
parametrization was used to smear the GEANT4-simulated detector 
response to produce resolution-corrected neutron scintillation-light 
yield spectra (see below).

\begin{figure} 
\begin{center}
\resizebox{13.5cm}{11cm}{\hspace{-1.5cm}\includegraphics{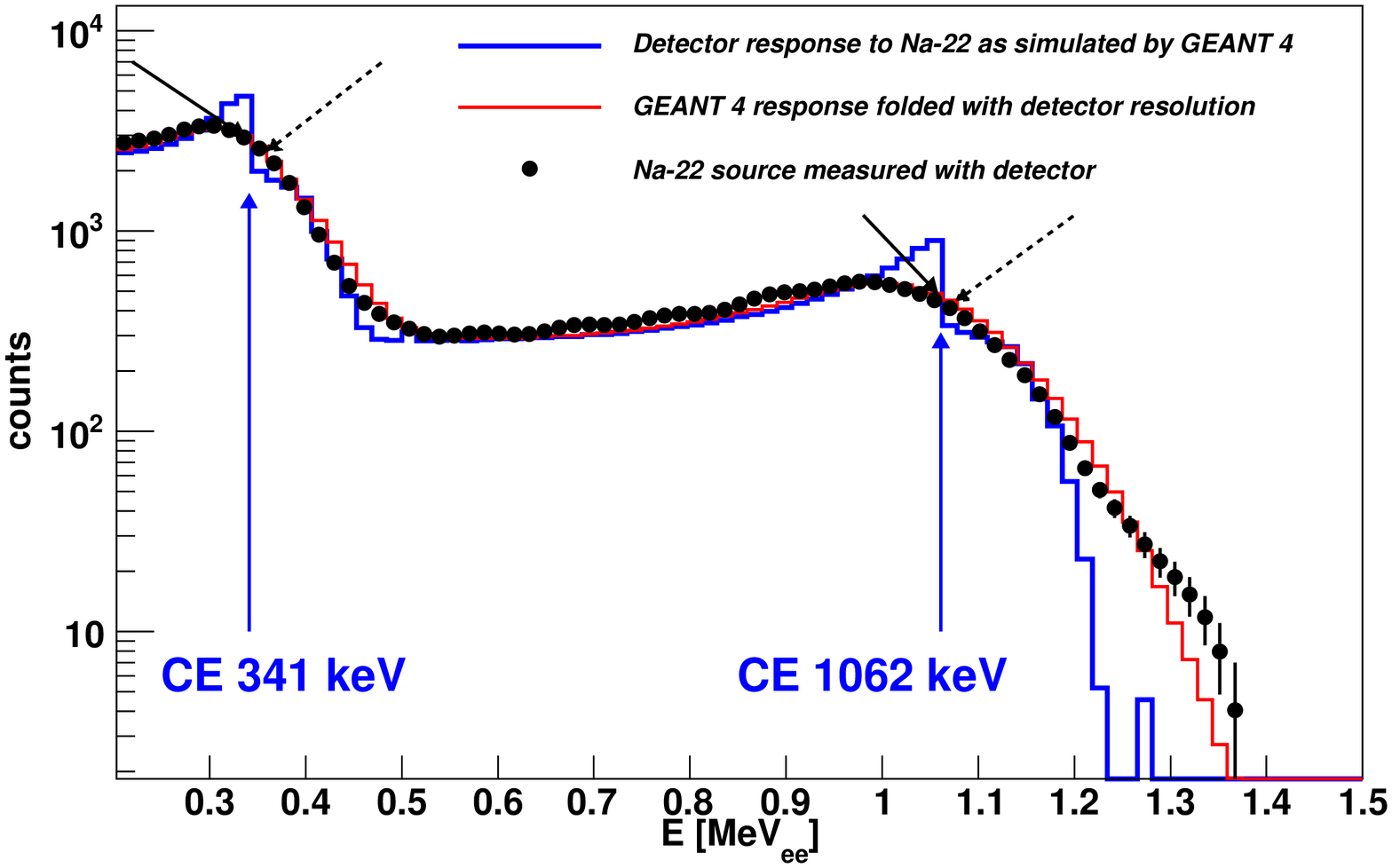}}
\caption{\label{figure:G4_calibration}
Simulated Compton scattered recoil-electron distributions (blue 
and red histograms) together with data (black dots) for $^{22}$Na 
as a function of the deposited energy / scintillation-light yield 
in MeV$_{ee}$. The blue histogram is the basic energy-deposition 
simulation which excludes energy resolution but clearly illustrates 
Compton edges at 341~keV$_{ee}$ and 1062~keV$_{ee}$. The red 
histogram is the simulation including resolution effects. The 
black dots result from the subtraction of non-source-related 
background from the measured data. The angled solid black arrows 
indicate the locations of the Compton edges determined using the 
method of Knox and Miller~\cite{knox72}, while the angled dashed 
black arrows indicate the locations of the Compton edges determined 
using the method of Flynn~et~al.~\cite{flynn64}. (For interpretation 
of the references to color in this figure caption, the reader is 
referred to the web version of this article.)
}
\end{center}
\end{figure}

Figure~\ref{figure:G4_calibration} shows a representative comparison 
between the GEANT4 simulation of the response of the detector to the 
gamma-rays coming from a $^{22}$Na source and background-subtracted 
data obtained with a $^{22}$Na source. The blue histogram corresponds 
to the basic simulation of the deposited energy and does not include 
resolution effects but clearly illustrates the Compton edges at 
341~keV$_{ee}$ and 1062~keV$_{ee}$. The red histogram corresponds to 
the simulation including the non-linear parametrization of the
energy-resolution effects detailed above. The black dots result from 
the subtraction of non-source-related background from the measured 
data. This included cosmic-ray background and experiment-hall 
background. The intensity of the cosmic-ray background was addressed 
with an energy-dependent exponential function, while the room 
background was addressed by identifying dominant gamma-rays present 
in data taken simultaneously with a HPGe detector. This background 
was then simulated as a combination of the dominant gamma-rays -- 
specifically, from $^{40}$K (1460~keV), $^{208}$Tl (2614~keV, 
583~keV, and 510~keV using the branching ratio 100:85:23) and 
511~keV positron
annihilation\footnote{
Unfortunately, a ``source-free" data set was not available.
}. 
The overall agreement between the 
measured data and simulation is very good. We attribute the very 
small variations between 0.5~MeV$_{ee}$ and 1.0~MeV$_{ee}$ to room 
background which we did not address. The enhanced strength at 1.4~MeV 
may be due to the simultaneous detection of both the gamma-rays 
emitted by $^{22}$Na.

We compared the results of our simulations to the Compton-edge
prescriptions suggested by Flynn~et~al.~\cite{flynn64} (dashed black 
arrows in Fig.~\ref{figure:G4_calibration}) and Knox and 
Miller~\cite{knox72} (solid black arrows in 
Fig.~\ref{figure:G4_calibration}). When the Flynn~et~al. approach 
was taken, we found it to overpredict systematically the locations 
of the Compton edges by more than 10\%. When the Knox and Miller 
approach was taken, we found it to underpredict systematically the 
locations of the Compton edges by less than 3\%.

\section{Results}
\label{section:results}

\begin{figure} 
\begin{center}
\resizebox{14cm}{14.5cm}{\hspace{-2.5cm}\includegraphics{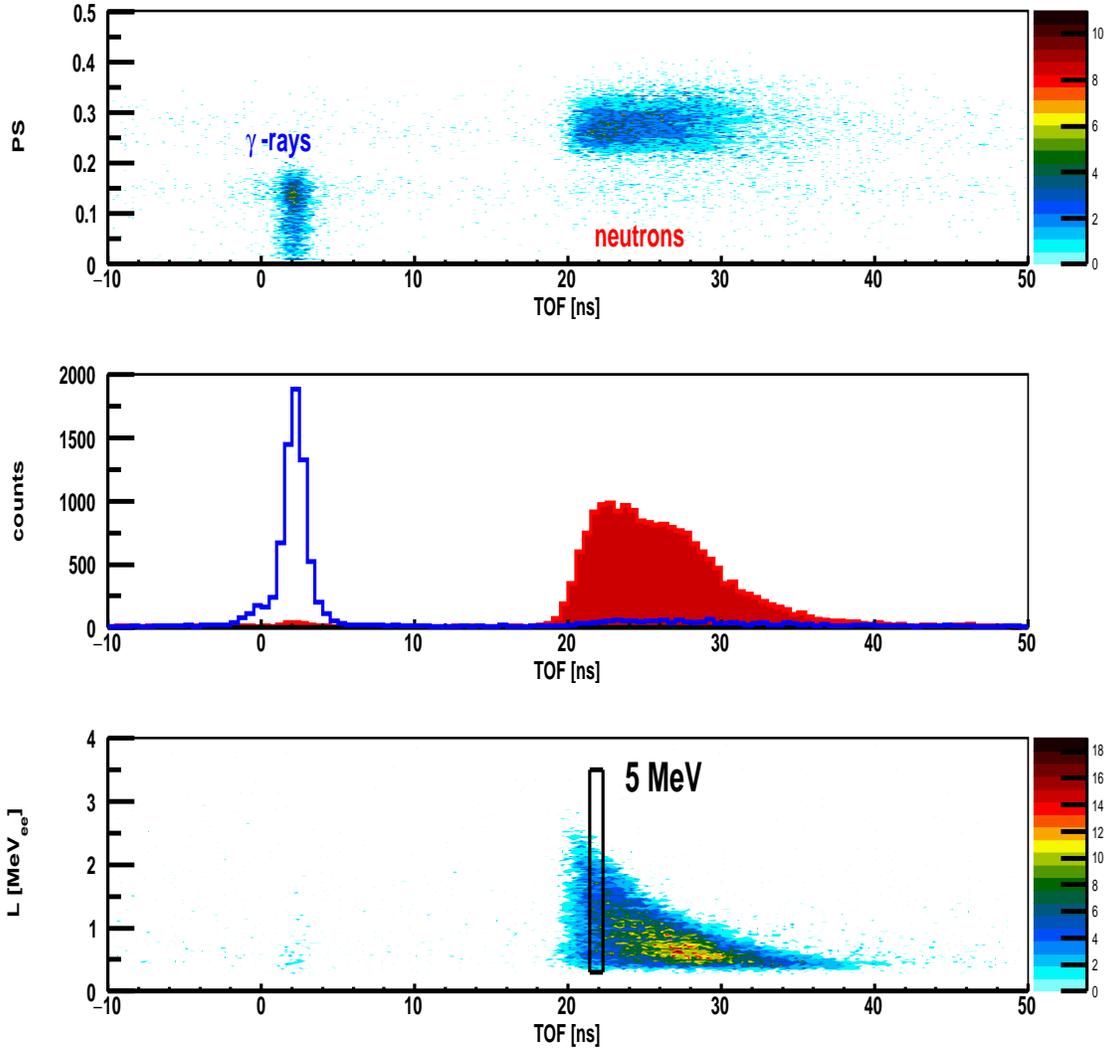}}
\caption{\label{figure:tof_light_yield}
Time-of-flight (TOF) distributions obtained for a neutron-drift 
distance of 0.675~m. Top panel: pulse shape (PS) plotted against TOF.  
Middle panel: projection of the data from the top panel onto the TOF 
axis. A PS~$=$~0.19 cut has been applied to separate gamma-rays and 
neutrons. The unshaded blue peak corresponds PS~$<$~0.19 while the 
shaded red distribution corresponds PS~$>$~0.19. Bottom panel: 
scintillation-light yield (L) plotted against TOF for PS~$>$~0.19. 
The cut to select neutrons with energy (5.0~$\pm$~0.1)~MeV 
(TOF~$\sim$~22~ns) is indicated with a black box. (For interpretation 
of the references to color in this figure caption, the reader is 
referred to the web version of this article.)
}
\end{center}
\end{figure}

As previously mentioned, gamma-ray scintillations in NE-213 are 
generally fast (10s of ns decay times) while neutron scintillations 
are much slower (100s of ns decay times). The type of radiation 
incident upon the NE-213 scintillator may thus be identified by 
examining the time structure of the scintillation pulses. We used 
the standard ``tail-to-total" 
method~\cite{jhingan08,lavagno10,pawelczak13}. With this method, 
the difference in the integrated charge produced by the 
scintillation-light pulses in the LG and SG QDCs was normalized to 
the integrated charge produced by the scintillation-light pulse in 
the LG QDC according to
\begin{equation}
PS = (LG - SG) / LG.
\end{equation}
Figure~\ref{figure:tof_light_yield} presents TOF distributions 
acquired when the NE-213 reference detector started the TOF TDC and 
the YAP(Ce) detector stopped the TOF TDC. The top two panels have been 
presented and discussed in detail in Ref.~\cite{scherzinger15} and 
are included here for completeness. The top panel illustrates that
the separation between gamma-rays (recoiling electrons) and neutrons 
(recoiling protons) was excellent. In the middle panel, the data 
from the top panel have been projected onto the TOF axis. The 
$\gamma$-flash and fast-neutron distributions are clearly identified. 
The very low level of background consists of random events which 
included cosmic rays, room background, Am/Be neutrons not correlated 
with a 4.44~MeV gamma-ray, and Am/Be neutrons where the 4.44~MeV 
gamma-ray was missed due to YAP inefficiency or geometry. In the 
bottom panel, our previously detailed calibration was applied to the 
data and the resulting scintillation-light yield is displayed for 
the SG QDC. It is this scintillation-light yield which we now proceed 
to analyze in detail.

The neutron scintillation-light yield (due to recoiling protons) was 
determined by converting from TOF to neutron kinetic energy, binning 
in widths of 0.2~MeV, and filling the corresponding energy-calibrated 
SG and LG QDC spectra. $T_{0}$ was determined from the location of the 
gamma-flash in the TOF spectra using the speed of light and 
measurements of the distances between the YAP(Ce) detector, the NE-213 
detector, and the Am/Be source. The neutron path length employed in 
this measurement was 2.420~m. Based upon our 1.8~ns gamma-flash and 
the detector thickness of 6.2~cm, we determined our energy resolution 
to be 4\% at 2~MeV and 5\% at 4~MeV. If the TOF bin width is 
sufficiently small and there was no smearing due to energy-resolution 
effects, each of these spectra would demonstrate a sharp cutoff 
corresponding to the neutron transferring all of its energy to the 
recoiling proton. In our detector, resolution effects smeared this 
maximum-transfer edge. Further, as in the case for locating the 
Compton edge for the energy calibration of organic scintillators with 
gamma-ray sources, there is no single prescription for relating the 
maximum proton energy to the resolution-smeared maximum-transfer edge. 
Thus, for each energy bin, we have investigated three 
edge-determination prescriptions:
\begin{enumerate}
\item{As suggested by Naqvi~et~al.~\cite{naqvi94}, a Gaussian function 
was fitted to the high-energy edge of the recoil-proton energy 
distribution and the maximum-transfer edge was taken to be the 
half-height (HH) position.}
\item{As suggested by Kornilov~et~al.~\cite{kornilov09}, the location 
of the most energetic minimum in the first derivative (FD) of the 
recoil-proton energy distribution was associated with the 
maximum-transfer edge.} 
\item{The maximum-transfer edge was taken as the turning point (TP) 
of the Gaussian function fitted to the resolution-smeared edge. Note 
that if the fitted Gaussian function described the resolution-smeared 
maximum-transfer edge perfectly, then the location of its TP is by 
definition identical to the minimum in the first derivative of the 
recoil-proton energy distribution.}
\end{enumerate}

In each investigation, the non-linear correspondence between 
recoiling electron ($E_{e}$) and recoiling proton ($E_{p}$) 
scintillation light-yield has been represented in two ways:
\begin{equation}
E_{e}=L_{0}\frac{E_{p}^{2}}{E_{p}+L_{1}}\label{eq:1_eq4_in_Kornilov09}
\end{equation}
\begin{equation}
E_{e}=C\{0.83E_{p}-2.82[1-\exp(-0.25E_{p}^{0.93})]\}\label{eq:2_from_Cecil79}
\end{equation}
Eq.~(\ref{eq:1_eq4_in_Kornilov09}) is the same as Eq.~(4) given in 
Ref.~\cite{kornilov09}, where $L_{0}$ and $L_{1}$ are adjustable 
parameters, and Eq.~(\ref{eq:2_from_Cecil79}) is from 
Ref.~\cite{cecil79}, where $C$ is an adjustable parameter.

\begin{figure} 
\begin{center}
\resizebox{13.5cm}{11cm}{\hspace{-1.5cm}\includegraphics{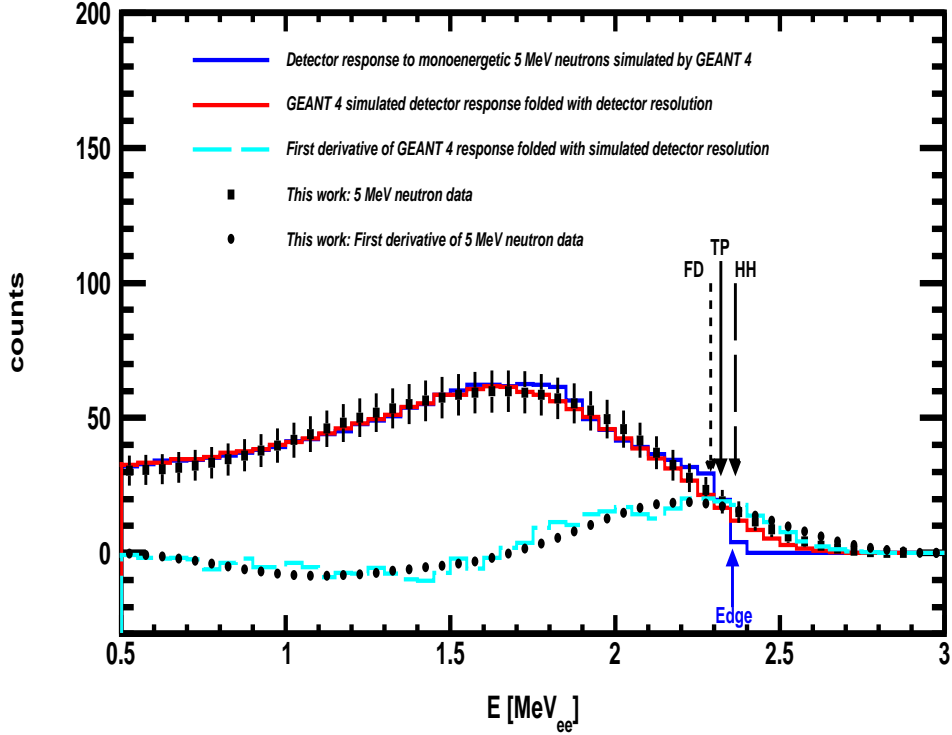}}
\caption{\label{figure:maximum_transfer_edge}
Simulated and measured neutron scintillation-light yield for 
(5.0~$\pm$~0.1)~MeV neutrons. The solid blue histogram shows the 
GEANT4-simulated 
detector response without resolution effects and the solid red 
histogram shows the GEANT4-simulated detector response folded with 
the measured detector resolution. The simulated maximum-transfer 
edge is clearly indicated. The dashed cyan histogram corresponds 
to the first derivative of the red histogram. Filled black squares 
correspond to measured data (statistical uncertainties are shown)
and filled black circles correspond to the first derivative of 
this measured distribution. The downward arrows point to the 
locations of the 5~MeV maximum-transfer edge according to the FD 
(short dash), TP (solid), and HH (long dash) prescriptions. (For 
interpretation of the references to color in this figure caption, 
the reader is referred to the web version of this article.)}
\end{center}
\end{figure}

\begin{figure} 
\begin{center}
\resizebox{13.5cm}{15cm}{\hspace{-1.5cm}\includegraphics{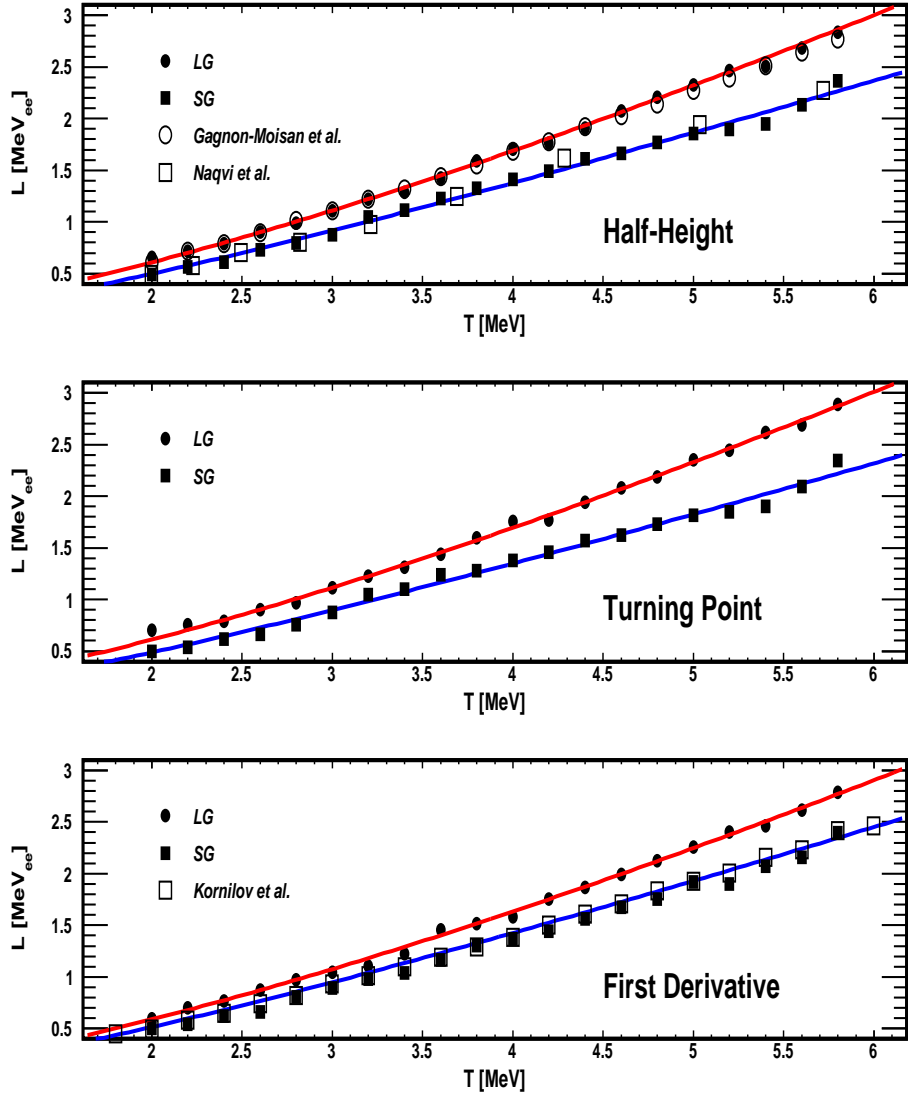}}
\caption{\label{figure:results}
LG (filled black circles) and SG (filled black squares) neutron 
scintillation-light yield (due to recoiling protons) as a function 
of neutron kinetic energy for different maximum-transfer edge 
determinations together with the data of 
Gagnon-Moisan~et~al.~\cite{gagnon12} (top panel, open circles), 
Naqvi~et~al.~\cite{{naqvi94}} (top panel, open squares), and 
Kornilov~et~al.~\cite{kornilov09} (bottom panel, open squares). 
The red curves are Eq.~(\ref{eq:1_eq4_in_Kornilov09}) fitted to 
the LG distributions while the blue curves are 
Eq.~(\ref{eq:2_from_Cecil79}) fitted to the SG distributions. 
(For interpretation of the references to color in this figure 
caption, the reader is referred to the web version of this 
article.)
}
\end{center}
\end{figure}

Figure~\ref{figure:maximum_transfer_edge} compares the 
GEANT4-simulated and measured neutron scintillation-light yield 
(due to recoiling protons) in the LG QDC for neutrons having 
(5.0~$\pm$~0.1)~MeV
kinetic energy (TOF $\sim$ 22~ns). In the GEANT4 simulation, 
the light-yield parametrization presented in 
Eq.~(\ref{eq:1_eq4_in_Kornilov09}) (see below for a discussion of 
light-yield parametrizations for NE-213) has been employed, where 
parameter $L_{1}$ was fixed at the Kornilov~et~al.~\cite{kornilov09} 
value of 2.47. The light-yield scaling (parameter 
$L_{0}$) was first based on the HH method for positioning the 
maximum-transfer edge (see the red curve at 5~MeV in the top panel 
of Fig.~\ref{figure:results}). 
The simulated detector response is shown without resolution effects, 
and the location of the maximum-transfer edge may be observed at 
about 2.36~MeV$_{ee}$. As the degree of smearing of the simulated 
detector response due to energy-resolution effects affects the 
location of the maximum-transfer edge in the simulated detector 
response predicted by the various prescriptions, the non-linear 
energy-dependent parametrization of the detector resolution 
measured for gamma-ray energies between 0.5~MeV$_{ee}$ and 
4.0~MeV$_{ee}$ that we employed in our calibration efforts was 
again employed to accurately smear the simulated detector response. 
The simulated detector response with resolution effects is also 
shown.
The arrows indicate the locations of the maximum-transfer edge in 
the data according to the FD (2.29~MeV$_{ee}$), TP (2.32~MeV$_{ee}$), 
and HH (2.36~MeV$_{ee}$) prescriptions. As expected, when 
the HH evaluation of the data is compared to the simulation with 
HH-based scaling, the agreement is essentially exact. The average 
location of the 5~MeV maximum-transfer edge is 2.33~MeV$_{ee}$, and 
all three predictions based upon the data agree to about 1\%. For 
these same 5~MeV neutrons, with the HH method for positioning the 
maximum-transfer edge fixed, we then varied the light-yield scaling 
in the GEANT4 simulation to the values obtained using the TP and FD 
prescriptions (see the red curves at 5~MeV in the middle and bottom 
panels of Fig.~\ref{figure:results}). In all three cases, the 
GEANT4 simulations were very close to the data up to 2~MeV$_{ee}$, 
with the FD and TP results lying at most 5\% and 3\% respectively 
above the HH results. Above 2~MeV$_{ee}$, comparison was difficult
due to a combination of lack of statistics and resolution effects.
For simulated (4.0~$\pm$~0.1)~MeV neutrons, light-yield scaling 
factors derived from all three methods resulted in a constant 8\% 
overestimation of the location of the maximum-transfer edges 
extracted from the data according to the procedure described above. 
At (3.0~$\pm$~0.1)~MeV the 
difference between the simulation-predicted and data-extracted edge 
location for the HH prescription remained at an 8\% overestimation, 
while for the TP and FD prescriptions, the discrepancy increased 
to 12\% and 18\%, respectively. At lower energies, the degree of 
non-linearity of the recoil-proton scintillation-light yield 
increases and this may account for the increasing discrepancy. 
Note that the use of Eq.~(\ref{eq:2_from_Cecil79}) gives very 
similar results from 3 to 5~MeV.

Fig.~\ref{figure:results} shows neutron scintillation-light yield 
(due to recoiling protons) data, as a function of neutron kinetic 
energy, for the three different maximum-transfer edge 
determinations together with existing results. The statistical 
uncertainties in our data are smaller than the point size. 
Gagnon-Moisan~et~al.~\cite{gagnon12} used a PS digitizer and 
employed a gamma-ray energy calibration similar to that of 
Ref.~\cite{knox72}. The tail-to-total method was used in their 
analysis together with the HH prescription for determining the 
maximum-transfer edge. Their data agree well with our 
corresponding LG results and we note that the method they used to 
integrate the total charge produced by the scintillation light is 
very similar to that employed here. Naqvi~et~al.~\cite{naqvi94} 
used ADCs (which we believe were peak-sensing) and employed the 
gamma-ray energy calibration suggested in Ref.~\cite{knox72}. The 
HH prescription was used for determining the maximum-transfer edge. 
Their data agree well with our corresponding SG results, but it is 
not possible to determine how much of the scintillation-light 
pulse was integrated from Ref.~\cite{naqvi94}. 
Kornilov~et~al.~\cite{kornilov09} used a charge-sensitive preamp 
together with an Ortec 460 delay-line amplifier and peak-sensing 
ADCs. Again, it is difficult to quantify how much of the 
scintillation pulse was integrated when the light yield was 
measured. ``Standard" (unspecified) gamma-ray energy calibrations 
were employed and the FD prescription was used to determine the 
maximum-transfer edge. Again their data agree well with our 
corresponding SG results.

Thus, the results from Ref.~\cite{naqvi94} and 
Ref.~\cite{kornilov09}, which used similar measurement techniques, 
are both in good agreement with our SG results. This could 
indicate that in these works, the entire charge associated with 
the scintillation was not integrated. On the other hand, real 
differences in the behavior of different samples of NE-213 are 
entirely possible and were observed in Ref.~\cite{kornilov09}.

In Fig.~\ref{figure:results}, the red curves shown in each panel 
display the scintillation-light yield parametrization described by 
Eq.~(\ref{eq:2_from_Cecil79}) fitted to the present LG data, while 
the blue curves display the scintillation-light yield parametrization 
described by Eq.~(\ref{eq:1_eq4_in_Kornilov09}) fitted to the SG data. 
In each case, the overall scale of the fitting function was allowed 
to float (see below). Note that when the fitted functions employed 
were interchanged (that is, when Eq.~(\ref{eq:2_from_Cecil79}) was 
fitted to our SG data and Eq.~(\ref{eq:1_eq4_in_Kornilov09}) was 
fitted to our LG data) the quality of fit was as good. This is not 
surprising as the scaled parametrizations differ by only about 3\% 
over this energy range.

Table~\ref{table:scale_factors} presents the scale factors 
(parameter $L_{0}$ of Eq.~(\ref{eq:1_eq4_in_Kornilov09}) and 
parameter $C$ of Eq.~(\ref{eq:2_from_Cecil79})) obtained by 
fitting parametrizations to the recoil-proton 
scintillation-light yield obtained with LG and SG. For 
Eq.~(\ref{eq:1_eq4_in_Kornilov09}), parameter $L_{1}$ was fixed 
at a value 2.47. In each case, the HH, TP, and FD prescriptions 
for determining the location of the maximum-transfer edge have 
been employed, and 20 data points between 2 and 6~MeV were 
considered.

\begin{table}
\begin{tabular}{|c|c|c|c|c|c|}
\hline 
Data & Edge & $L_{0}$ (from Eq.~(2)) & $\chi^{2}/$d.o.f & $C$ (from Eq.~(3)) & $\chi^{2}$/d.o.f.\tabularnewline
\hline 
\hline 
LG & HH & $0.704\pm0.006$ & 0.86 & $1.056\pm0.009$ & 1.20\tabularnewline
\hline 
SG & HH & $0.555\pm0.005$ & 0.98 & $0.828\pm0.007$ & 1.10\tabularnewline
\hline 
SG/LG & HH & $0.789\pm0.010$ &  & $0.784\pm0.009$ & \tabularnewline
\hline 
LG & TP & $0.702\pm0.006$ & 1.37 & $1.044\pm0.010$ & 1.41\tabularnewline
\hline 
SG & TP & $0.543\pm0.005$ & 1.27 & $0.810\pm0.007$ & 1.29\tabularnewline
\hline 
SG/LG & TP & $0.774\pm0.010$ &  & $0.776\pm0.010$ & \tabularnewline
\hline 
LG & FD & $0.689\pm0.005$ & 1.05 & $1.037\pm0.005$ & 0.74\tabularnewline
\hline 
SG & FD & $0.539\pm0.005$ & 1.19 & $0.813\pm0.005$ & 0.82\tabularnewline
\hline 
SG/LG & FD & $0.783\pm0.010$ &  & $0.784\pm0.006$ & \tabularnewline
\hline 
\end{tabular}
\caption{\label{table:scale_factors}
Scale factors $L_{0}$ and $C$ from fits of Eq.~(2) and Eq.~(3)
to the present LG and SG data together with ratios. ``Edge" denotes 
the method used to determine the maximum-energy edge of the 
recoil-proton scintillation-light yield.}
\end{table}

There is little to choose between the $\chi^{2}$/d.o.f. values 
which are all close to 1. Further, the ratios of scale parameters 
for the SG and LG data do not differ significantly between any of 
the edge-determination prescriptions or between the use of 
Eq.~(\ref{eq:1_eq4_in_Kornilov09}) or Eq.~(\ref{eq:2_from_Cecil79})
for the correspondence between the recoiling electron and recoiling 
proton scintillation-light yield. Comparing the present LG values 
of $L_{0}$ with those presented in Ref.~\cite{kornilov09} where a 
similar value of $L_{1}$ was used, our values are a factor 
$\sim 1.2$ higher. Thus, compared to Ref.~\cite{kornilov09}, we have 
collected a factor 1.2 more recoil-proton scintillation.  On the 
other hand, from the values of $C$ presented in 
Table~\ref{table:scale_factors} which are only a few percent above 1, 
it can be seen that our results are quite similar to those presented 
in Ref.~\cite{cecil79} and close to those presented in 
Ref.~\cite{gagnon12}.

\section{Summary and Discussion}
\label{section:summary}

We have reported a detailed mapping of the response of a NE-213 
detector to neutrons from 2--6 MeV emitted by a lead-shielded Am/Be 
source and subsequently tagged by time-correlated gamma-ray emission. 
Neutron/gamma pulse shape discrimination was performed using the 
gated tail-to-total QDC method, with charge-integration periods set 
to 35~ns and 475~ns. The electron-energy calibration was performed 
using standard gamma-ray sources and two prescriptions for locating 
the corresponding Compton edges were examined. The results were 
compared to GEANT4 simulations which considered both
energy-resolution effects and backgrounds. The Compton-edge 
prescriptions of Knox and Miller~\cite{knox72} and 
Flynn~et~al.~\cite{flynn64} differ by more than 10\% when applied to 
our data. The present GEANT4 simulations suggest that the former 
underpredicts the actual edge position by $\sim3$\%, while the latter 
overpredicts by $\sim10$\%. Consequently, we used the prescription of 
Knox and Miller~\cite{knox72} scaled up by a factor 1.03.

The present neutron-tagging technique provided a continuous, 
polychromatic, energy-tagged neutron beam from 2--6~MeV. Neutron 
kinetic energy was determined by measuring the neutron TOF relative 
to the prompt 4.44~MeV gamma-ray associated with the 
$\alpha+^{9}$Be$\rightarrow n+^{12}$C$^{*}$ process. Using this 
information, recoil-proton scintillation-light yields were determined 
as a function of neutron kinetic energy. Three different prescriptions 
were employed for identifying the maximum energy-transfer edge of the 
recoiling protons in accumulated neutron scintillation-light spectra. 
Two parametrizations (Eq.~(\ref{eq:1_eq4_in_Kornilov09}), 
Eq.~(\ref{eq:2_from_Cecil79})) of the recoil-proton 
scintillation-light yield were investigated. Simple scaling factors 
allowed for variations in the neutron scintillation-light yield, and 
after scaling, either parametrization fitted our LG and SG data equally 
well.

GEANT4 was also used to study the effects of the three prescriptions 
for the determination of the recoil-proton edge in the neutron 
scintillation light-yield spectra. For a fixed light-yield 
parametrization, we varied the prescription between HH, TP, and FD 
(both with and without energy-resolution effects) for 3, 4, and 5~MeV 
neutrons. At 5~MeV, simulation and analysis agreed for all 
prescriptions at the 1\% level. At 4~MeV, all three GEANT4 predictions 
for the maximum-transfer edge overestimated the location of the 
maximum-transfer edge extracted from the data by 8\%. At 3~MeV, the 
difference between the edge locations extracted from the simulation 
and data for the HH prescription remained at 8\%, while for the TP 
and FD prescriptions, the difference increased to 12\% and 18\%, 
respectively. A possible cause of the disprecancy is an incomplete 
consideration of increasing quenching of the scintillation as 
$dE/dx$ increases along the track of the recoiling proton. This will 
be investigated in future work. Nevertheless, the HH method produced 
the best results for our detector over our energy range.

The present results indicate that for recoiling protons in the present 
energy range, $\sim78$\% (see Table~\ref{table:scale_factors}) of the 
total integrated scintillation intensity (integration period 475~ns) 
is contained with the first 35~ns of the signal. Comparing the total 
light yield (LG) to previous measurements, the present results are in 
good agreement with those of Gagnon-Moisan~et~al.~\cite{gagnon12} and 
within a few percent of those of Cecil~et~al.~\cite{cecil79}, the 
latter of which being used to estimate recoil-proton 
scintillation-light output in the absence of a calibration. The present 
LG results are higher by a factor $\sim1.2$ compared to those of 
Kornilov et~al.~\cite{kornilov09} and Naqvi et~al.~\cite{naqvi94}. 
These previous measurements yield results which are actually close to 
our SG results (integration period 35~ns), but it is impossible to say 
if this discrepancy is due to a difference in effective integration 
times, as the pulse-processing method was different. At least part of 
the disprecancy could be due to real differences in the response of the 
liquid scintillator. Factors such as concentration of the active 
scintillant/fluorescent materials in the base solvent and the presence 
of dissolved oxygen will affect the relative recoiling 
proton-to-electron scintillation-light yields. Indeed, it would seem 
that a dedicated measurement of the recoil-proton scintillation-light 
yield must be made on a case-by-case basis to obtain the best accuracy 
in precision neutron measurements.

The present measurements have been made at a new neutron test facility 
recently installed at Lund University~\cite{scherzinger15}. This 
facility is being used to measure the characteristics of neutron 
detectors as part of the program to build the European Spallation 
Source. Development and extension of this facility is ongoing with a 
view to precisely determining the response of many materials to 
neutrons ranging in energies from fast to thermal.

\section*{Acknowledgements}
\label{acknowledgements}

We thank the Photonuclear Group at the MAX IV Laboratory for providing 
access to their experimental hall and Am/Be source. We acknowledge the 
support of the UK Science and Technology Facilities Council (Grant nos. 
STFC 57071/1 and STFC 50727/1) and the European Union Horizon 2020
BrightnESS Project, Proposal ID 676548.

\newpage

\bibliographystyle{elsarticle-num}

\end{document}